\def\BibTeX{{\rm B\kern-.05em{\sc i\kern-.025em b}\kern-.08em
    T\kern-.1667em\lower.7ex\hbox{E}\kern-.125emX}}
\documentclass[10pt,conference]{IEEEtran}

\usepackage[sort&compress,numbers]{natbib}
\usepackage{multirow}
\usepackage{enumitem}
\usepackage{balance}
\usepackage[most]{tcolorbox}
\usepackage{graphicx}
\usepackage{amsmath}
\usepackage{hyperref}
\usepackage{pifont}
\usepackage{colortbl}
\usepackage{amsfonts}

\usepackage{booktabs}

\usepackage{varwidth} 
\tcbuselibrary{breakable} 
\tcbuselibrary{skins} 
\usepackage{algorithm}
\usepackage{algorithmic}
\usepackage{subcaption}
\usepackage{soul}

\definecolor{lightgray}{gray}{0.9}
\definecolor{lightgreen}{rgb}{0.9, 1, 0.9} 
\definecolor{lightred}{rgb}{1, 0.9, 0.9} 

\usepackage[framemethod=tikz]{mdframed}
\mdfdefinestyle{customstyle}{
  linecolor=gray!100,
  linewidth=3pt,
  innerleftmargin=3pt,
  topline=false,
  rightline=false,
  bottomline=false,
  leftline=true,
  innerrightmargin=3pt,
  innertopmargin=3pt,
  innerbottommargin=3pt,
  backgroundcolor=gray!15
}
\newtcolorbox{mycodebox}[1][]{
    sharp corners,
    boxrule=0.8pt,
    colframe=gray!80!black,
    colback=white,
    boxsep=3pt,
    left=0pt, right=0pt, top=0pt, bottom=0pt,
    #1 
}
\definecolor{darkblue}{rgb}{0, 0, 0.5}

\newenvironment{custommdframed}
  {\begin{mdframed}[style=customstyle]}
  {\end{mdframed}}
 
\begin{document}

\title{\textit{AutoPLC}: Generating Vendor-Aware Structured Text for Programmable Logic Controllers}

\author{%
  \IEEEauthorblockN{%
    \parbox{0.6\textwidth}{\centering 
      Donghao Yang\textsuperscript{a},
      Aolang Wu\textsuperscript{a},
      Tianyi Zhang\textsuperscript{a},
      Li Zhang,
      Xiaoli Lian\textsuperscript{b},
      Fang Liu\textsuperscript{b},
      Yuming Ren,
      Jiaji Tian
    }%
  }%
  \IEEEauthorblockA{%
    SKLSDE, Beihang University, China\\
    Email: \{yangdonghao,aolangwoo,tianyiz,lily,\\
    lianxiaoli,fangliu,yumingren,tianjiaji\}@buaa.edu.cn\\
    {\footnotesize\textsuperscript{a}Equal contribution,
    \textsuperscript{b}Co-corresponding authors}%
  }%
  \and
  \IEEEauthorblockN{Xiaoyin Che}%
  \IEEEauthorblockA{%
    Siemens AG\\
    Email: xiaoyin.che@siemens.com%
  }%
}%
\maketitle%

\begin{abstract}
Among the programming languages for Programmable Logic Controllers (PLCs), Structured Text (ST) is widely adopted for industrial automation due to its expressiveness and flexibility. However, major vendors implement ST with proprietary extensions and hardware-specific libraries - Siemens' SCL and CODESYS' ST each differ in syntax and functionality. This fragmentation forces engineers to relearn implementation details across platforms, creating substantial productivity barriers.
To address this challenge, we developed AutoPLC, a framework capable of automatically generating vendor-aware ST code directly from natural language requirements. Our solution begins by building two essential knowledge sources tailored to each vendor's specifications: a structured API library containing platform-exclusive functions, and an annotated case database that captures real-world implementation experience. Building on these foundations, we created a four-stage generation process that combines step-wise planning (enhanced with a lightweight natural language state machine support for control logic), contextual case retrieval using LLM-based reranking, API recommendation guided by industrial data, and dynamic validation through direct interaction with vendor IDEs.
Implemented for Siemens TIA Portal and the CODESYS platform, AutoPLC achieves 90\%+ compilation success on our 914-task benchmark (covering general-purpose and process control functions), outperforming all selected baselines, at an average cost of only \$0.13 per task. Experienced PLC engineers positively assessed the practical utility of the generated code, including cases that failed compilation. We open-source our framework at \textcolor{darkblue}{\url{https://github.com/cangkui/AutoPLC}}.

\end{abstract}


\section{Introduction}
\label{sec:intro}

Programmable Logic Controllers (PLCs) are the cornerstone of Industrial Control Systems (ICSs) within the broader field of Industrial Automation. They play a pivotal role in the operation and
management of critical infrastructure across diverse sectors such as energy, manufacturing, and transportation.  The significance of PLCs is further evidenced by their growing market projection,
with forecasts anticipating a rise to USD 12.20 billion by 2024 and USD 15.12 billion by 2029, marking a robust Compound Annual Growth Rate (CAGR) of 4.37\% from 2024 to 2029 ~\cite{PLCMarket,PLCMarket-MultiCountry}.

PLCs are programmed predominantly using languages defined by the IEC 61131-3 standard~\cite{IEC61131}. Among these, Structured Text (ST) is the only high-level, block-structured textual language, with syntax resembling Pascal and C. Due to its high-level features, engineers often use ST to implement data processing, communication protocols, and complex control algorithms. In scenarios where graphical languages like Function Block Diagram fall short in expressiveness, ST becomes the preferred choice~\cite{PLComplexTask}.

As industrial control systems become increasingly complex, interoperability among heterogeneous devices from different vendors becomes essential\cite{MO2023172}. However, vendors typically implement ST with proprietary extensions to IEC 61131-3, including custom syntax, specialized functions, and firmware-specific execution semantics. Notable examples include Siemens' SCL, Rockwell's ST, and CODESYS' ST, each featuring distinct syntax and functionality. This vendor-specific fragmentation imposes a significant learning burden on engineers and highlights the critical need for automated, vendor-aware code generation in industrial control development.

Large Language Models (LLMs) have made remarkable progress in code generation, though primarily for mainstream languages like Java and Python~\cite{yin2018mining, austin2021program, wu2024versicodeversioncontrollablecodegeneration, jimenez2024swebench}. This focus stems from the abundance of high-quality public code available on platforms like GitHub. 
However, applying LLMs to ST generation introduces unique challenges. High-quality public ST datasets are scarce, hindering effective training and domain adaptation. Directly transferring frameworks from general-purpose languages often leads to code hallucination~\cite{tianCodeHaluCodeHallucinations2024a}\cite{haag2024trainingllmsgeneratingiec}, such as generating Pascal-style control structures. Additionally, LLMs often struggle with vendor-specific functions, particularly those introduced in recent firmware versions, resulting in type mismatches and unsupported function calls.

Despite these challenges, both academia and industry have begun exploring LLM-based automation for ST programming. For instance, LLM4PLC~\cite{fakih2024llm4plc} fine-tuned LLMs on limited public dataset, achieving preliminary ST code generation. Koziolek et al.~\cite{koziolekLLMbasedRetrievalAugmentedControl2024} proposed retrieval-augmented generation (RAG) to integrate relevant function blocks into the output, while Agents4PLC~\cite{liu2024agents4plc} introduced a multi-agent framework combining RAG with a feedback-refinement loop to improve accuracy.
However, these pioneering explorations generally lack practicality in current industrial environments. The fundamental reasons are twofold:

\begin{itemize}[leftmargin=1em]

    \item \textbf{Limited utilization of vendor knowledge.} Accurate ST code generation requires a precise understanding of vendor-specific syntax and fundamental functions. However, these resources are typically fragmented across technical documentation, lacking structured formats compatible with RAG systems. Although Agents4PLC~\cite{liu2024agents4plc} attempts to curate reusable code snippets and domain knowledge, its opaque knowledge base construction methodology prevents effective reuse. In addition, existing approaches~\cite{koziolekLLMbasedRetrievalAugmentedControl2024, liu2024agents4plc}  retrieve vendor functions via embedding similarity with control narratives, failing to align high-level requirements and low-level function semantics, resulting in irrelevant retrievals. 

    \item \textbf{Lack of interaction with industrial programming environments.} Most approaches use open-source compilers (\textit{e.g.}, MATIEC~\cite{matiec}) without interacting with actual PLC programming platforms~\cite{fakih2024llm4plc,liu2024agents4plc}. Moreover, existing evaluation benchmarks are small in scale.~\citet{fakih2024llm4plc} evaluate their approach on a small subset of the OSCAT IEC 61131-3 library (i.e., 40 samples)~\cite{OSCAT}, while~\citet{haag2024trainingllmsgeneratingiec} use the generic APPS Python benchmark, translating it into ST via GPT-4, which may not reflect real-world PLC patterns.~\citet{liu2024agents4plc} develop a custom benchmark tailored to ST-specific features, but only 23 programming tasks are publicly released. These limitations make it difficult to reliably assess the automation potential in real-world industrial scenarios.
\end{itemize}

To address these challenges, we propose \emph{AutoPLC}, a novel framework for generating vendor-specific ST code. Our approach combines two specialized knowledge bases: (1) \emph{Rq2ST} - a case library of requirement-plan-code triplets enriched with vendor APIs, and (2) \emph{APILib} - a comprehensive function library integrating both IEC 61131-3 standards and vendor extensions.
The framework implements an intelligent four-stage workflow: planning (using lightweight NL-defined finite state machines for control logic), case retrieving, API recommendation (combining align their semantics with the intent of each planned step, I/O type matching, and case-based reasoning), and an iterative generate-improve process that actively interfaces with vendor platforms.

To evaluate the effectiveness of \emph{AutoPLC} in generating ST code, we constructed a benchmark dataset of 914 tasks based on open-source resources. The dataset covers two representative ST programming scenarios:
1) General-purpose tasks, such as date processing, string manipulation, file I/O, and numerical computation;
2) Process control tasks, such as robotic arms and conveyor systems.
It includes programming tasks for two vendors: the open-source CODESYS' ST language and Siemens' industrial Structured Control Language (SCL) dialect~\cite{CODESYSGroup2024}.

We implement AutoPLC and conduct experiments on two mainstream vendor platforms, Siemens' TIA Portal~\cite{SIMATICSTEP7a} and 3S-Smart's CODESYS~\cite{CODESYSGroup2024}. Results demonstrate that AutoPLC outperforms state-of-the-art ST generation methods and general-purpose code models, achieving over 90\% compilation pass rate on generating the two ST variants, with the average cost of \$0.13 per task. In comparison, the advanced general model Claude-3.5 only achieves 27\% on TIA Portal and 58\% on CODESYS. On the Agents4PLC benchmark with 23 tasks, AutoPLC achieves a formal verification pass rate of 78\%. To further assess its practical utility, we invite domain experts to review the generated code. The experts provide positive feedback on its effectiveness and applicability in real-world industrial scenarios.

In summary, our main contributions are as follows.

\begin{itemize}[leftmargin=1em]
\item \textbf{Methodology:}
We propose the \textit{AutoPLC} framework, which includes a vendor knowledge base construction approach and a LLM-based pipeline for generating ST code from natural language requirements.

\item \textbf{Benchmark:}
We construct a benchmark dataset of 914 requirement–code pairs, covering general-purpose and process control tasks in both CODESYS' ST and Siemens' SCL.

\item \textbf{Evaluation:} We conduct a comprehensive evaluation of \textit{AutoPLC}, and the results demonstrate that it outperforms state-of-the-art methods,
exhibits promising generalizability across ST variants.

\item \textbf{Open-source Package:}
We release a replication package containing the benchmark, source code, and experimental results. The \emph{AutoPLC} system interfaces with and is evaluated on two industrial platforms: Siemens TIA Portal and CODESYS. 
\end{itemize}

\section{Background and Problem Statement}
\label{sec:background}

\subsection{Background}

\begin{figure*}[!ht]
    \centering
    \setlength{\abovecaptionskip}{0.1cm}
    \includegraphics[height=7.3cm, keepaspectratio]{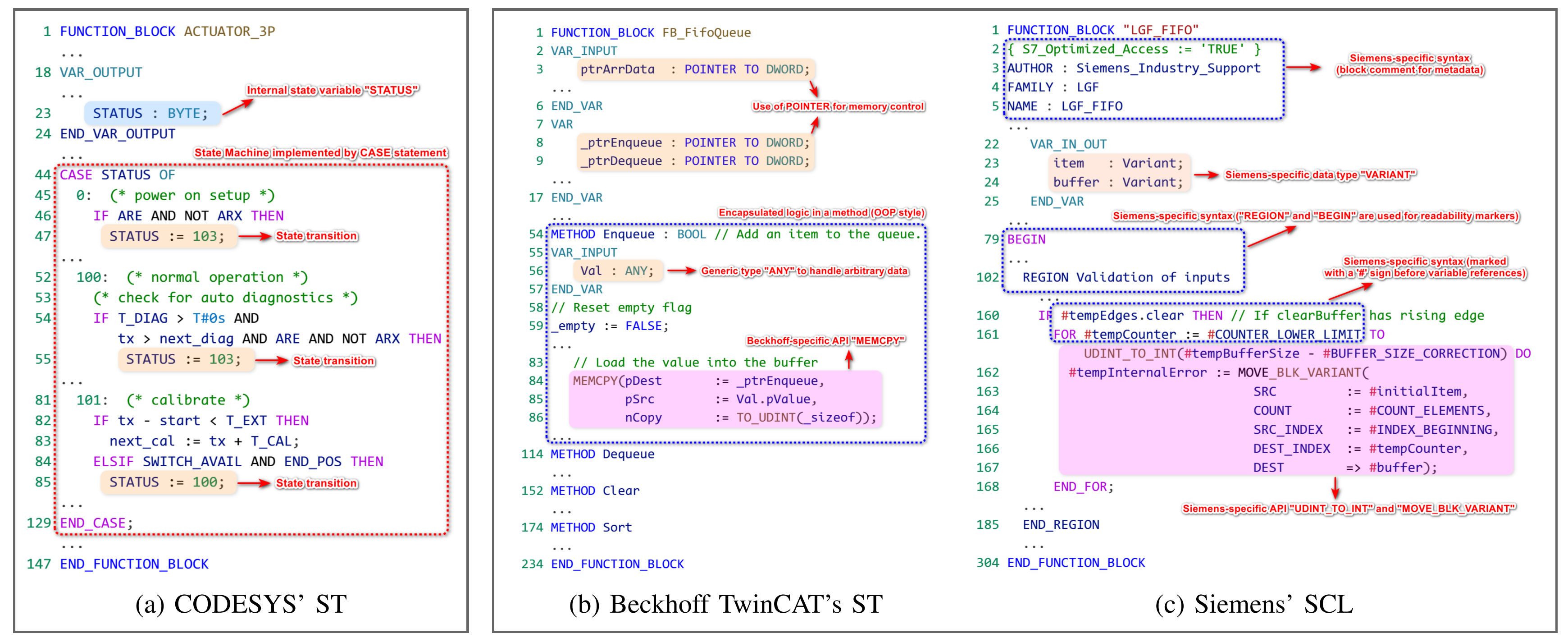}
    \caption{Example code segments implemented by different vendors. (a) is ACTUATOR, (b) and (c) are both FIFO.}
    \label{fig:casecodes}
    \vspace{-0.4cm}
\end{figure*}

A PLC project typically consists of multiple units, such as FUNCTION\_BLOCKs with internal state and stateless FUNCTIONs, each encapsulating specific control logic and interacting with one another through defined input/output interfaces, covering both general-purpose logic (e.g., arithmetic, comparison, timing) and device-specific control (e.g., motors, sensors, actuators).

Fig.\ref{fig:sub_a} demonstrates an actuator control logic implementation using a finite state machine in CODESYS' ST. The solution maintains internal state variables and updates outputs according to system timing constraints. Figs. \ref{fig:sub_b}-\ref{fig:sub_c} compare the syntactic differences when implementing a FIFO (First-In-First-Out) FUNCTION\_BLOCK across two vendor platforms:

\begin{itemize}[leftmargin=1em]

    \item Fig.\ref{fig:sub_b} shows the implementation for Beckhoff TwinCAT-based PLCs, which support object-oriented programming. Developers can use \texttt{POINTER} types and \texttt{ANY} generics to handle arbitrary data types, and encapsulate logic in methods such as \texttt{Enqueue}, offering both high-level abstraction and low-level memory control similar to general-purpose languages.
    
    \item Fig.\ref{fig:sub_c} presents the counterpart for Siemens S7-1200/1500 PLCs, which adopt a procedural style. It relies on proprietary data type \texttt{Variant} to handle arbitrary data, and uses dedicated instructions like \texttt{MOVE\_BLK\_VARIANT} to manipulate this type.
    
\end{itemize}

Clearly, developing units for PLC project requires a thorough understanding of vendor ST's syntax, libraries, and programming paradigms. This increases the complexity of cross-vendor development.

\subsection{Problem Statement}

Given $\mathit{Task} = \langle \mathit{Req}, \mathit{IO}, \mathit{Type}, \mathit{Vendor\_Target} \rangle$ as task description, where \texttt{Req} is a natural language description of the desired functionality, $\texttt{IO} = \{(t_i, v_i, s_i, d_i)\}_{i=1}^n$ is a list of I/O variables with type $t_i$, name $v_i$, description $s_i$, and direction $d_i \in \{\texttt{IN}, \texttt{OUT}, \texttt{INOUT}\}$, $\texttt{Type}$ specifies whether to generate a FUNCTION\_BLOCK or a FUNCTION, and \texttt{Vendor\_Target} specifies the target PLC platform. The goal is to generate a set $\mathcal{C} = \{c_1, c_2, \dots, c_k\}$ of vendor-aware ST units such that each $c_i$ is expected to be compilable on the specified platform or require only minimal manual adjustment. 
 
\section{Approach: AutoPLC}
\label{sec:approach}

\begin{figure*}[!ht]
    \centering
    \setlength{\abovecaptionskip}{0.1cm}
    \includegraphics[trim = {2cm 12cm 1cm 7cm}, clip, width=1.0\textwidth]{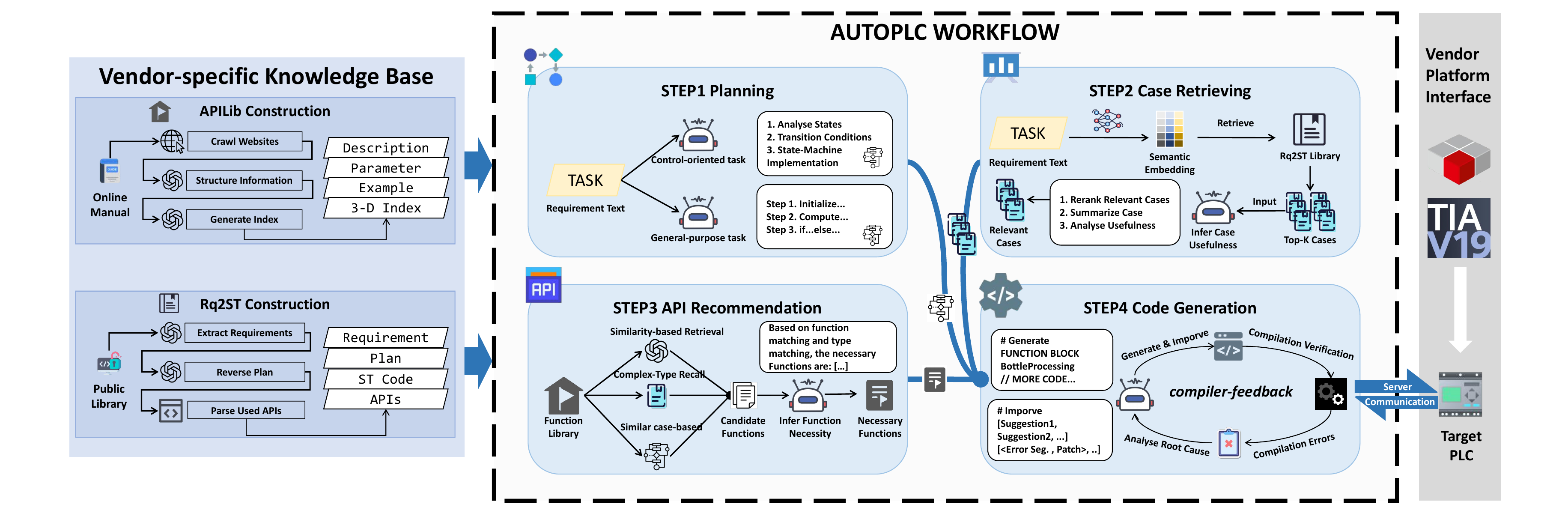}
    \caption{Framework of AutoPLC.}
    \label{fig:framework}
    \vspace{-0.4cm}
\end{figure*}

\emph{AutoPLC} provides an end-to-end solution for generating industrial ST code from natural language requirements. The framework comprises two fundamental components: a vendor-specific knowledge infrastructure and a four-stage LLM-based workflow, as shown in Fig.\ref{fig:framework}.

\emph{AutoPLC} operates through two specialized knowledge bases. \emph{APILib} serves as a comprehensive repository of vendor fundamental APIs, complete with functional descriptions and usage examples, while \emph{Rq2ST} maintains a collection of proven requirement-to-code implementations. The generation proceeds through four key stages: Planning, Case Retrieving from \emph{Rq2ST}, API Recommendation from \emph{APILib}, and final Code Generation based on interaction with platforms.

Thus, extending \emph{AutoPLC} to new ST variants is straightforward - simply update the underlying \emph{APILib} and \emph{Rq2ST} components, and integrate the compilation service of target PLC vendor platform during the synthesis phase. This modular approach preserves the core generation workflow while adapting to different vendor environments.

\subsection{Vendor-specific Knowledge Base Construction}

\emph{APILib} assists correct usage of vendor-specific functions(APIs), while \emph{Rq2ST} enables \emph{AutoPLC} to learn from proven implementation experience.

\emph{APILib} provides structured metadata for the fundamental APIs (usually hardware-aware), systematically collected from vendor documentation. Each API entry includes complete specifications (name, description, parameters) and practical examples. To support API Recommendation in Stage 3, we use GPT-4o to generate a three-dimensional index for each API, comprising: (1) a functional summary, (2) typical usage scenarios, and (3) ST-specific keywords. This index serves as a compact semantic representation for efficient retrieval. Formally, each API is defined as $\mathit{API = \langle Name, Desc, Params, Examples, Index} \rangle$, where $\mathsf{Index} = \langle S_{\text{summary}}, S_{\text{scenario}}, K_{\text{keywords}} \rangle$ contains the generated summary, scenarios, and keyword set.

Our successful case library \emph{Rq2ST} draws from multiple sources, including vendor-specific reusable function sets (like Siemens' SCL libraries or CODESYS' instruction set) and validated industrial code repositories. 
Each case is represented as a quadruple $\langle \mathit{Task}, \mathit{Plan}, \mathit{Code}, \mathit{APIs} \rangle$ containing the task description, 
the design plan, the implementation, and the fundamental APIs. While $Task$ and $Code$ are obtained directly, the high-level $Plan$ is derived by GPT-4o through code analysis, and the $APIs$ are identified by matching the ST code against \emph{APILib}. 

\subsection{Stage 1: Planning}

Planning prior to code implementation serves as a critical translation mechanism, converting high-level natural language task descriptions into executable \textbf{step-by-step} specifications. It has been empirically validated as an effective approach for bridging abstract requirements with concrete implementations, ultimately enhancing the quality of generated code~\cite{jiang2024self-planning,dong2024self-collaboration}.

State-of-the-art planning for code generation is typically implemented using advanced LLMs~\cite{jiang2024self-planning,dong2024self-collaboration,MapCoder}, and some studies propose pseudocode-based planning to enhance code generation accuracy~\cite{li2023structured}, we identify two critical limitations in applying these techniques directly to our task:

\begin{itemize}[leftmargin=1em]
    \item \emph{Overly generic plans fail to capture control logic specifics.} Public LLMs, trained predominantly on general-purpose tasks, tend to produce plans that inadequately represent the state transitions fundamental to process control tasks.
    
   \item \emph{Pseudocode-formatted plans introduce syntactic noise in ST generation.} According to our observation, when plans are expressed in pseudocode, the code-generation LLM may directly replicate pseudocode constructs due to its inability to properly distinguish them from valid ST syntax. 
\end{itemize}

Our approach maintains the use of advanced LLMs for planning while introducing key adaptations. First, we implement task-type classification (process-control vs. general-purpose) through targeted prompting. For process-control tasks, our Planner directs the LLM to: (1) identify system states and their transitions, and (2) formulate a natural-language state machine representation, with execution planning reduced to state machine traversal. For general-purpose tasks, we follow conventional linear planning methodologies~\cite{li2023structured,zhang2023planning} that outline procedural logic through sequential computational steps.

We deliberately employ natural-language plans instead of pseudocode to avoid syntactic contamination. 
This abstraction preserves the LLM's flexibility in selecting appropriate ST implementations while preventing direct syntax carryover from intermediate planning representations.

\subsection{Stage 2: Case Retrieving}

Despite the structured task decomposition from the planning stage, generating compliant ST code remains non-trivial due to the dialect variations across vendor implementations and the fact that ST is under-represented in the pre-training corpora of LLMs.
While domain adaptation through fine-tuning represents a common approach for LLM specialization~\cite{LLMCG,guo2024deepseek,Nijkamp2023codegen}, 
the per-vendor cost of continuously curating large-scale ST corpora quickly outweighs the benefits, making fine-tuning economically unattractive for our multi-vendor setting.
Recent advances in RAG~\cite{zhaoRetrievalAugmentedGenerationAIGenerated2024,RAGsurvey} 
offer a more viable solution by capitalizing on LLMs' demonstrated in-context learning capabilities~\cite{brown2020gpt3,dong2024surveyincontextlearning}
: the compact, high-quality cases in Req2ST are fetched on demand, letting the LLM learn in context without ever retraining.

Our retrieval process begins by selecting 5 candidate cases based on embedding similarity between the task description and case's metadata, leveraging Zhipu AI's retrieval service\footnote{\url{https://bigmodel.cn/dev/howuse/retrieval}}. Next, we employ an LLM to rerank these candidates according to their potential contribution to the current task, ultimately choosing the three most beneficial examples for final inclusion. We then dynamically inject their requirements, plans, and ST code as few-shot examples in Stage 4's structured prompt. Although these retrieved cases inherently include API usage information, we do not directly inject these APIs as few-shot demonstrations. Instead, the relevant APIs undergo processing and refinement in Stage 3 before being systematically integrated into the generation prompt.

\subsection{Stage 3: API Recommendation}

The goal of this stage is to recommend potentially useful APIs for the ST generation. This approach has proven effective in improving the performance of code-generation LLMs, particularly for less familiar languages~\cite{10554837}. The challenge of API usage is further compounded by vendor-specific extensions to the standard (e.g. Siemens adds hundreds of proprietary low-level library functions, named Extended Instructions). 
Consequently, general-purpose LLMs often fail to accurately identify the appropriate hardware-aware functions across diverse scenarios, especially given the concrete dialect and limited resources of ST.

We adopt a two-step retrieval and filtering strategy to balance recall and precision. Recall is prioritized, as providing a broad candidate set enables the LLM to make informed selections. Without sufficient recall, the LLM lacks viable options, even if it can assess relevance accurately~\cite{10.1145/3729347}. To further enhance recall, we collect candidate APIs from three dimensions during retrieval: similarity-based retrieval, parameter-based augmentation, and similar-case-based augmentation. Denote the APIs collected from these dimensions as  $\Lambda_{similarity}$, $\Lambda_{par}$, and $\Lambda_{simCase}$. The final API set for this stage is the union of these three sets.

The set $\Lambda_{similarity}$ is obtained by using each step in the planning process from Stage 1 as a query to retrieve candidate APIs, calculating relevance against a 3-dimensional index of each API. We employ Okapi BM25~\cite{Robertson2009bm25}, a method widely used in RAG for code-generation tasks~\cite{zhang2023syntax,liao2024a3codgen}. The set 
$\Lambda_{par}$ includes potential APIs that share input or output types with the ongoing task, focusing specifically on ST-specific complex types (e.g., structured data types like \texttt{DTL}). The set $\Lambda_{simCase}$ is constructed by retrieving APIs from the top three relevant cases identified in Stage 2.

Next, we apply LLM-based filtering to refine the final set. The filtering needs to consider both functional relevance (whether the API's purpose matches the task) and interface compatibility (whether parameter types align), which cannot be fully determined through keyword matching alone. Thus, we prompt an LLM to evaluate and select the most suitable functions. To maintain prompt efficiency, we package the indexed fields of up to 15 candidate functions into each prompt. While no filtering is perfect, research shows that LLMs can effectively assess the suitability of retrieved content, making this approach both practical and empirically validated~\cite{10.1145/3729347,saad-falcon-etal-2023-udapdr}.

\subsection{Stage 4: Code Generation}

In this stage, we dynamically construct a structured prompt by combining the task description with contextual elements: step-by-step plan, potentially useful APIs, and requirement-plan-code few-shot examples.
This rich prompt enables the LLM to generate high-quality ST code. To ensure syntactic correctness, we implement a feedback-driven modification loop that iteratively improves initially generated code using compiler error analysis (called the Self-improvement mechanism).

While open-source ST validation tools like MATIEC~\cite{matiec} and IECChecker~\cite{iecchecker_2024} exist, they lack support for the diverse and evolving ST dialects. Our solution integrates vendor platform's services, e.g., TIA Openness (Siemens TIA Portal), providing access to authentic compilation diagnostics directly within our generation workflow.

When fixing errors in the generated code, we follow a practical way informed by engineers experience: we tackle \textit{declaration errors} first before moving to \textit{implementation errors}, since unresolved declarations tend to create a flood of cascading issues (e.g. “undefined identifier”). Our repair process works by having the LLM carefully inspect each error, figure out potential fixes, and then produce snippet-replacement patches. These targeted patches get applied to the problematic snippets, allowing for precise corrections without having to regenerate the entire code. We keep refining this way until the code compiles cleanly or we hit our iteration limit.

\section{Evaluation}
\label{sec:evaluation}

\subsection{Research Questions}

\noindent \textbf{RQ1 (Effectiveness Evaluation):} What is the effectiveness of \emph{AutoPLC} in generating ST across vendors?

\noindent \textbf{RQ2 (Ablation Study)}: How do AutoPLC's Planning, Case Retrieving, API Recommendation, and the Self-improvement mechanism for code generation affect overall performance?

\noindent \textbf{RQ3 (Manual Evaluation):} How do developers perceive the quality of code generated by \emph{AutoPLC}?

\subsection{Datasets and Metrics}

\subsubsection{\textbf{Datasets}}
\label{subsubsec:datasets}

Due to the sensitivity of industrial applications, publicly available ST code is scarce. Existing studies, such as Fakih et al.~\cite{fakih2024llm4plc}, constructed benchmarks based on public ST libraries but did not release their datasets, limiting transparency and reproducibility. In this study, we build our benchmark datasets by carefully curating 914 cases from three authoritative sources representing two major industrial vendors' ST variants (CODESYS' ST and Siemens' SCL):

\begin{enumerate} [leftmargin=1em]
    \item CODESYS Implementation of OSCAT Library (718 cases)~\cite{OSCAT}: A widely-adopted, community-maintained open-source library containing general ST functions, network utilities, and building automation components. 
    \item Siemens LGF Library (151 cases)~\cite{LibraryGeneralFunctions}: An industrial-standard library optimized for Siemens automation systems, offering comprehensive general-purpose functions.
    \item Siemens Competition Dataset (45 cases)~\cite{GenerativeAIApplication}: Real-world industrial programming tasks from Siemens featuring process control task with practical constraints.
\end{enumerate}

For each function in these library documentations, we extract the description and input/output parameters as task description, and the corresponding code implementation as the ground truth. We further analyze the code complexity based on the vendor's official implementation. Table~\ref{tab:Rq2ST_statistics} provides detailed statistics including case counts, lines of code (LOC), and input/output variables.
 
Additionally, we evaluate AutoPLC on the Agents4PLC Benchmark~\cite{liu2024agents4plc}, which includes 23 PLC programming tasks spanning diverse scenarios.

\begin{table}[!htbp]
\centering
\scriptsize
 \setlength{\abovecaptionskip}{0.1cm}
\caption{Statistics of Our Benchmark.}
\label{tab:Rq2ST_statistics}
\resizebox{\linewidth}{!}{
\begin{tabular}{lccccc}
    \toprule
    \multirow{2}{*}{\textbf{Dataset}} & \multirow{2}{*}{\textbf{\#Cases}} & \multicolumn{2}{c}{\textbf{Lines of Code}} & \multicolumn{2}{c}{\textbf{Input/Output Variables}}  \\
    \cmidrule(lr){3-4}
    \cmidrule(lr){5-6}
            &       & Average & Median & Average & Median \\
    \midrule
    OSCAT & 718   & 34.40 & 20 & 4.39 & 2 \\
    LGF         & 151   & 91.46 & 69 & 5.50 & 4 \\
    Competition & 45    & 84.68 & 76 & 6.12 & 6 \\
    \bottomrule
\end{tabular}
}
\vspace{-0.3cm}
\end{table}

\subsubsection{\textbf{Metrics}}
\label{subsubsec:metrics}

Due to the lack of standardized ST benchmarks~\cite{fakih2024llm4plc}, we adopt dataset-specific evaluation metrics. For our constructed datasets (i.e., LGF, Competition, and OSCAT), we report the \emph{Compilation Pass Rate (Pass Rate)} and \emph{Average Error Count (Avg. Errors)}, based on actual compilation results from CODESYS and Siemens TIA Portal. We omit formal verification due to unavailable public specifications and prohibitive construction costs for large-scale datasets~\cite{NIANG2020103328}. Following~\citet{koziolekAutomatedControlLogic2024a}'s OSCAT framework, we implement rigorous validation by first verifying test cases against the ground truth implementations (retaining only tasks with $\geq$ 5 passing cases), then manually executing test suites in CODESYS with full success required for verification.

For the 23 tasks in the Agents4PLC benchmark, we follow its official protocol~\cite{liu2024agents4plc}, using \texttt{rusty} for syntax checking and \texttt{plcverif} for formal verification. 

\subsection{Implementations of AutoPLC}

Our implementation generates ST code for two major industrial platforms.
\begin{itemize}[leftmargin=1em]
    \item \emph{Siemens' SCL}: We construct the Rq2ST knowledge base using the Competition and LGF datasets (Section~\ref{subsubsec:datasets}), augmented with an APILib derived from Siemens' official documentation\footnote{\url{https://docs.tia.siemens.cloud/p/plc-programming-with-simatic-s7}}. During feedback-based code generation (Stage 4), the system interfaces with TIA Portal V19 on S7-1200/1500 PLCs (firmware v2.9) to validate compilation.
    \item \emph{CODESYS' ST}: The Rq2ST knowledge base integrates the OSCAT dataset and APILib from CODESYS' official documentation. Compilation is performed using CODESYS V3.5 SP20 Patch 5.
\end{itemize}
 
To prevent data leakage in code generation (Stage 2 Case Retrieving), we implement a strict filtering mechanism: for each task $t$, any case $e$ is excluded from retrieval (i.e., omitted from \emph{Rq2ST}) if their textual names exhibit mutual substring containment (case-insensitive). While some cases implementing similar logic to the task may remain, we argue that this reflects real-world project conditions—and is precisely why RAG is beneficial.

In Stage 4, we set the maximum iteration count to 3. For the backbone LLM, we selected Claude-3.5-Sonnet~\cite{Claude} due to its robust code generation capabilities. Our evaluation shows that \emph{AutoPLC} achieves practical efficiency, with a cost of \$0.13 per task (matching LLM4PLC) and a processing time of 52 seconds, making it viable for real-world deployment.

\subsection{Baselines}\label{sec:baseline}

We compare AutoPLC against state-of-the-art baselines, including LLM4PLC~\cite{fakih2024llm4plc}, Agents4PLC~\cite{liu2024agents4plc}, MapCoder~\cite{MapCoder}, and several widely-used general LLMs.

For LLM4PLC, we adapt the framework to a fully autonomous setting by 
removing the manual interventions while retaining its iterative feedback from MATIEC~\cite{matiec}(syntax checking) checking and NuXMV~\cite{nuXmv}(formal verification)
, following its original SMV-based verification strategy.
Since the original paper does not specify iteration counts, we standardize this to three iterations, consistent with AutoPLC. To ensure fairness, we adopt the latest GPT-4o as the backbone, as GPT-4 was reported but not version-specified in the original paper~\cite{fakih2024llm4plc}.

Due to the lack of the released implementation details for Agents4PLC, we are unable to reproduce its results, but we include a comparison against the reported performance in their paper on their released benchmark (Agents4PLC Benchmark)~\cite{liu2024agents4plc}. For MapCoder, we use the source code downloaded from the official GitHub directly, and set its internal \texttt{self.LANGUAGE} configuration to ST/SCL, while keeping the other settings unchanged. 
We employ the same backbone LLM (Claude-3.5-Sonnet) as used in our \emph{AutoPLC} system.

Additionally, we include six general LLMs: DeepSeek-Chat~\cite{guo2024deepseek}, Qwen2.5-Coder-Instruct (7B)~\cite{qwenwebsite}, GPT-4o~\cite{GPT4}, GLM-4-Plus~\cite{glm4pluswebsite}, Claude-3.5-Sonnet~\cite{Claude}, and Llama-3.1-Instruct (8B)~\cite{llama3website}. These models are widely used as coding assistants and cover a diverse range of architectures and capabilities.\footnote{Due to API availability, we use DeepSeek-V3 for the Agents4PLC benchmark and DeepSeek-V2.5 for others.}
To ensure consistency, all models are evaluated using the same prompt templates, which are provided in the supplementary replication package.

\section{Results and Analysis}

\subsection{Addressing RQ1: Effectiveness Evaluation}

\begin{table*}[t]
\centering
\setlength{\abovecaptionskip}{0.1cm}
\small
\caption{Comparison results of \emph{AutoPLC} and baselines in generating ST and SCL code.}
\resizebox{1.0\textwidth}{!}{
\begin{tabular}{lcccccccc}
\toprule
\multirow{2}{*}{\textbf{Baselines}} 
& \multicolumn{2}{c}{\textbf{LGF (SCL)}} 
& \multicolumn{2}{c}{\textbf{Competition (SCL)}} 
& \multicolumn{2}{c}{\textbf{OSCAT (ST)}} 
& \multicolumn{2}{c}{\textbf{Agents4PLC (ST)}} \\
\cmidrule(lr){2-3} \cmidrule(lr){4-5} \cmidrule(lr){6-7} \cmidrule(lr){8-9}
 & \textbf{Pass Rate} & \textbf{Avg. Errors} 
 & \textbf{Pass Rate} & \textbf{Avg. Errors} 
 & \textbf{Pass Rate} & \textbf{Avg. Errors} 
 & \textbf{Pass Rate} & \textbf{Valid. Rate} \\
\midrule

\textbf{Llama-3.1-Instruct\footnotesize (8B)} & 0.66\% & 15.65 & 0.00\% & 22.84 & 5.98\% & 28.14 & 4.55\% & 0.00\% \\
\textbf{Qwen2.5-Coder-Instruct\footnotesize (7B)} & 5.96\% & 10.92 & 5.61\% & 9.56 & 18.08\% & 13.14 & 73.91\% & 17.39\%  \\
\textbf{GLM-4-Plus} & 6.62\% & 10.54 & 6.67\% & 9.36 & 35.05\% & 12.22 & 82.61\% & 30.43\%  \\
\textbf{GPT-4o} & 8.61\% & 18.76 & 20.00\% & 13.49 & 35.47\% & 8.39 & 82.61\% & 43.48\% \\
\textbf{DeepSeek-Chat} & 3.97\% & 14.53 & 22.22\% & 13.27 & 42.70\% & 7.68 & 82.61\% & 39.13\% \\
\textbf{Claude-3.5-Sonnet} & 15.23\% & 9.60 & 26.67\% & 6.53 & 57.72\% & 5.76 & 78.26\% & 52.17\% \\
\midrule
\textbf{LLM4PLC} & 7.28\% & 16.54 & 9.09\% & 21.80 & 30.74\% & 15.52 & 36.36\% & 4.55\% \\
\textbf{Agents4PLC\footnotesize (paper)} & - & - & - & - & - & - & 100.00\% & 60.87\% \\
\textbf{MapCoder} & 11.26\% & 12.97 & 35.56\% & 18.34 & 56.96\% & 8.60 & 82.61\% & 56.52\% \\
\textbf{AutoPLC} & \textbf{92.72\%} & \textbf{1.14} & \textbf{91.11\%} & \textbf{1.27} & \textbf{92.90\%} & \textbf{1.58} & \textbf{100.00\%} & \textbf{78.30\%} \\
\bottomrule
\end{tabular}
}
\label{tab:RQ1}
\vspace{-0.3cm}
\end{table*}

Table \ref{tab:RQ1} presents the comparative results of \emph{AutoPLC} against all baselines across four datasets. \emph{AutoPLC} demonstrates consistent superiority over all baselines in both evaluation metrics, achieving compilation pass rates exceeding 90\% while maintaining fewer than 2 average errors across all datasets.

More observations can be made as follows.
\begin{itemize}[leftmargin=1em]
    \item \emph{Claude-3.5-Sonnet shows competitive performance} among all baselines, rivaling specialized PLC LLMs like LLM4PLC and MapCoder. As our approach's backbone, it ranks second-best across our three datasets. Nevertheless, \emph{AutoPLC} achieves remarkable improvements over Claude-3.5-Sonnet: pass rates increase by 60.95\% (OSCAT),  508.8\% (LGF), and  241.6\% (Competition), while average errors decrease by  72.6\%,  88.1\%, and  80.6\% respectively. On the Agents4PLC benchmark, \emph{AutoPLC} achieves 100\% compilation success and the highest validation pass rate, outperforming LLM4PLC and MapCoder by 1620.8\% and 38.3\% respectively.
    Notably, while MapCoder also uses Claude-3.5-Sonnet, it underperforms the native model in some cases 
    (e.g., on the LGF dataset, Claude surpasses MapCoder by 26.1\% in compilation pass rate and produces 35.1\% fewer errors).
    This contradicts the expected performance gains from multi-agent systems in general NLP tasks~\cite{MapCoder}, likely because MapCoder's example-recalling agent depends on model's internal knowledge that proves ineffective for scarce ST examples, precisely where our vendor-specific knowledge bases add value.
    
    \item \emph{LLM4PLC underperforms expectations}. Our implementation reveals that the base GPT-4o outperforms LLM4PLC in both ST and SCL generation. Unlike the original LLM4PLC paper, we employ fully automated verification via MATIEC (syntax checking) and NuXMV (formal verification). Our analysis identifies two key limitations: \ding{182} MATIEC sometimes misflags correct code as erroneous, providing misleading feedback; \ding{183} GPT-4o struggles to generate valid formal model and specifications for verification. These findings suggest that: (1) industrial vendor platforms should replace open-source MATIEC to improve reliability, and (2) human intervention remains essential for formal verification at this stage.

    \item \emph{Our approach shows robust performance across two variants, whereas all baselines demonstrate 3.03-10.75× better ST than SCL generation performance (compare the OSCAT and LGF datasets)}. Models show highest pass rates on OSCAT, moderate performance on Competition, and poorest results on LGF, a trend correlating with average LOC (see Table \ref{tab:Rq2ST_statistics}. Longer code usually indicates greater complexity). For error analysis (limited to LGF and Competition due to compiler differences), models consistently make more errors on LGF, aligning with pass rate findings. However, our approach does not show obvious difference between these two SCL datasets due to the usage of two knowledge bases and the feedback-based improvement during code generation.
    The baselines exhibit a pronounced performance gap between ST and SCL generation, as evidenced by OSCAT (ST benchmark) and LGF (SCL benchmark) evaluations, where ST achieves 3.03-10.75× higher pass rates. This disparity stems from two factors: \ding{182} OSCAT's alignment with ST's widespread industrial adoption (e.g., Siemens ecosystems) versus LGF's specialization for niche controllers, resulting in richer ST training data; and \ding{183} SCL's dependence on Siemens-specific extensions beyond IEC 61131-3, which reveal baseline models' knowledge limitations. Notably, \emph{AutoPLC} maintains robust performance across both language-benchmark pairs.
\end{itemize}

\textbf{Execution-based evaluation.} To evaluate functional correctness beyond compilation, we executed the generated test suites on AutoPLC and MapCoder—the latter being a specialized code-generation model with strong performance on general-language test-case accuracy \cite{MapCoder}. Following the methodology in Section \ref{subsubsec:metrics}, we selected the intersection of compilation-passing cases from both approaches. After filtering, we retained 71 tasks, each with approximately 10 test cases.

The results show that AutoPLC achieved full-suite success on 35 tasks, while MapCoder succeeded on 34—both close to 50\%. Across all 487 test cases, the two models passed ~76\%. This validates our approach, especially considering OSCAT’s 92.90\% compilation pass rate. Notably, MapCoder’s low compilation pass rate but high test case pass rate suggests that the primary challenge in ST generation lies in syntax compliance, a prerequisite for test execution.
All raw test results are available in the replication package.

\vspace{1mm}
\begin{custommdframed}
\textbf{Answer to RQ1:} \emph{AutoPLC} markedly outperforms state-of-the-art baselines across all datasets, with higher compilation pass rate and lower error counts.  
\end{custommdframed}
\vspace{-0.2cm}

\subsection{Addressing RQ2: Ablation Study}
\label{subsec:alabtion}

To rigorously evaluate \emph{AutoPLC}'s design, we conduct an ablation study by sequentially disabling Planning, Case Retrieving, API Recommendation, and Self-improvement mechanism in Stage 4. Notably, the vendor-specific knowledge bases (\emph{Rq2ST} and \emph{APILib}) serve as foundational infrastructure that cross-cut multiple Stages - particularly enabling Case Retrieving (Stage 2) and API Recommendation (Stage 3). Therefore, rather than ablating these shared knowledge bases directly, we assess their implicit impact through the ablation of the dependent algorithmic components.

\begin{table*}[t]
\centering
\setlength{\abovecaptionskip}{0.1cm}
\small
\caption{Ablation study results.}
\resizebox{1.0\textwidth}{!}{
\begin{tabular}{lcccccccc}
\toprule
\multirow{2}{*}{\textbf{Baselines}} 
& \multicolumn{2}{c}{\textbf{LGF (SCL)}} 
& \multicolumn{2}{c}{\textbf{Competition (SCL)}} 
& \multicolumn{2}{c}{\textbf{OSCAT (ST)}}
& \multicolumn{2}{c}{\textbf{Agents4PLC (ST)}} \\ 
\cmidrule(lr){2-3} \cmidrule(lr){4-5} \cmidrule(lr){6-7} \cmidrule(lr){8-9}
 & \textbf{Pass Rate} & \textbf{Avg. Errors} 
 & \textbf{Pass Rate} & \textbf{Avg. Errors} 
 & \textbf{Pass Rate} & \textbf{Avg. Errors}
 & \textbf{Pass Rate} & \textbf{Valid. Rate} \\
\midrule

\textbf{w/o Planning} 
& 91.39\%\tiny{\textcolor{red}{$\downarrow$1.43\%}} 
& 2.52\tiny{\textcolor{red}{$\uparrow$121.05\%}} 
& 86.67\%\tiny{\textcolor{red}{$\downarrow$4.87\%}} 
& 1.24\tiny{\textcolor{green}{$\downarrow$2.36\%}} 
& 87.13\%\tiny{\textcolor{red}{$\downarrow$6.22\%}} 
& 2.79\tiny{\textcolor{red}{$\uparrow$76.58\%}} 
& 95.65\%\tiny{\textcolor{red}{$\downarrow$4.35\%}} 
& 69.57\%\tiny{\textcolor{red}{$\downarrow$11.15\%}} \\

\textbf{w/o Case Retrieving} 
& 60.26\%\tiny{\textcolor{red}{$\downarrow$35.01\%}} 
& 5.08\tiny{\textcolor{red}{$\uparrow$345.61\%}} 
& 73.33\%\tiny{\textcolor{red}{$\downarrow$19.51\%}} 
& 3.09\tiny{\textcolor{red}{$\uparrow$143.31\%}} 
& 85.10\%\tiny{\textcolor{red}{$\downarrow$8.39\%}} 
& 1.93\tiny{\textcolor{red}{$\uparrow$22.15\%}} 
& 86.96\%\tiny{\textcolor{red}{$\downarrow$13.04\%}} 
& 43.48\%\tiny{\textcolor{red}{$\downarrow$44.47\%}} \\

\textbf{w/o API Recommendation} 
& 84.77\%\tiny{\textcolor{red}{$\downarrow$8.57\%}} 
& 2.82\tiny{\textcolor{red}{$\uparrow$147.37\%}} 
& 84.44\%\tiny{\textcolor{red}{$\downarrow$8.93\%}} 
& 2.75\tiny{\textcolor{red}{$\uparrow$116.54\%}} 
& 87.85\%\tiny{\textcolor{red}{$\downarrow$5.44\%}} 
& 1.93\tiny{\textcolor{red}{$\uparrow$22.15\%}} 
& 100.00\% 
& 73.91\%\tiny{\textcolor{red}{$\downarrow$5.61\%}} \\

\textbf{w/o Self-improvement} 
& 66.23\%\tiny{\textcolor{red}{$\downarrow$28.57\%}} 
& 7.02\tiny{\textcolor{red}{$\uparrow$515.79\%}} 
& 64.44\%\tiny{\textcolor{red}{$\downarrow$29.27\%}} 
& 1.57\tiny{\textcolor{red}{$\uparrow$23.62\%}} 
& 84.80\%\tiny{\textcolor{red}{$\downarrow$8.75\%}} 
& 2.89\tiny{\textcolor{red}{$\uparrow$82.91\%}} 
& 100.00\% 
& 69.57\%\tiny{\textcolor{red}{$\downarrow$11.15\%}} \\

\textbf{AutoPLC (Full)} 
& \textbf{92.72\%} & \textbf{1.14} 
& \textbf{91.11\%} & \textbf{1.27} 
& \textbf{92.90\%} & \textbf{1.58} 
& \textbf{100.00\%} & \textbf{78.30\%} \\ 
\bottomrule
\end{tabular}
}
\label{tab:ablation}
\vspace{-0.3cm}
\end{table*}

The outcomes are presented in Table~\ref{tab:ablation}. To quantify the extent of these variations, we include the change ratios accompanied by directional arrows and red figures to signify the degree of reduction or increase, respectively. The following observations can be made:

\begin{itemize}[leftmargin=1em]

    \item \emph{All modules contribute to the generation of ST and SCL code, as evidenced by increased Pass Rates and reduced Average Errors when incorporated.}

    \item \emph{The impact of the stages is more pronounced in the SCL datasets compared to the ST datasets.} 
    For example, removing the Case Retrieving causes an 8.39\% drop in the Pass Rate for OSCAT, but 35.01\% and 19.51\% for LGF and Competition, respectively, highlighting a 26.62\% gap between public libraries. This result reinforces the conjecture that general LLMs lack domain-specific expertise for SCL programming. As shown in Huang et al. (2024)~\cite{huang2024trustllmtrustworthinesslargelanguage}, incorporating external knowledge and feedback strategies can substantially boost performance.

    \item \emph{The case-retrieving and the self-improvement contribute substantially to performance gains.} 
    Removing the case-retrieving reduces the Pass Rate by up to 35.01\%, and removing the self-improvement leads to a decrease of up to 29.27\% in the Pass Rate. Notably, case-retrieving removal has a more severe impact on the LGF benchmark compared to Competition benchmark, likely because LGF tasks, focused on general-purpose functions, rely heavily on Siemens' SCL extensions. 
    For the Agents4PLC benchmark, excluding the case-retrieving critically degrades both Compilation Pass Rate and Validation Satisfied Rate, underscoring its role in supplying proven, domain-appropriate logic experience that help raise the overall reliability of control programs.

    \item \emph{The API-recommendation and the Planning stages provide complementary improvements and are still valuable.} 
    Although their individual impact is less dramatic than Case Retrieving and the Self-improvement, they contribute to steady performance gains across datasets, validating their role in assisting function selection and decompose requirement. 
    
\end{itemize}

\vspace{1mm}
\begin{custommdframed}
\textbf{Answer to RQ2:} The Case Retrieving and the Self-improvement drive notable gains, while the API Recommendation and the Planning provide complementary support.
\end{custommdframed}
\vspace{-0.2cm}

\subsection{Addressing RQ3: Manual Evaluation}

Till now, we have assessed quantitative measures by evaluating the compilation pass rate and the count of errors for each snippet of generated code. However, we aim to delve deeper into understanding \emph{to what extent can the generated code be practically useful in real engineering scenarios?} To explore this question, we have carried out a user study where participants are invited to manually evaluate aspects such as \emph{correctness, conformance to industry coding standards, modifiability, safety, and usefulness} for the generated code.

\begin{itemize}[leftmargin=1em]
    \item Correctness: On a scale from 1 to 5, to what degree the provided code fulfills the specified tasks.
    \item Conformance to Industry Coding Standards: Rated from 1 to 5, this measures how well our code aligns with industry practices and norms, including aspects like variable naming conventions, and structuring of branches and loops.
    \item Modifiability: On a scale from 1 to 5, how easily our code can be altered or expanded upon, particularly within the ever-changing industrial environment.
    \item Safety: This category, scored from 1 to 5, explores the extent to which our code could be executed safely within a real-world industrial setting.
    \item Usefulness: On a scale from 1 to 5, usefulness gauges the extent to which our code serves as a helpful reference for the evaluator's own coding work. 
\end{itemize}

The first four metrics are designed for cases that successfully pass compilation, while the final metric is specifically intended for cases that fail to compile. To guarantee the validity of the evaluations, participants were asked to provide explanations for their assigned scores.

We randomly assembled a collection of 30 cases from each of our datasets, amounting to a total of 90 cases. Each case includes a specified requirement and its associated code segment. In every individual benchmark, 20 cases result in successful compilation, while the remaining 10 lead to compilation failures.

We recruited five experts. Three are veteran automation electrical engineers with extensive PLC development experience, counting 8, 7, and 2.5 years respectively. And two are doctoral candidates with specializations in electrical industrial automation, each possessing three years of ST programming experience applied to diverse assignments and projects with varying degrees of complexity. These participants were engaged through public job posting platforms and had no prior knowledge of the authors. They were each responsible for assessing between 10 to 15 cases based on their time constraints. 

\begin{figure}[!ht]
    \centering
    \setlength{\abovecaptionskip}{0.1cm}
    \includegraphics[width=0.45\textwidth]{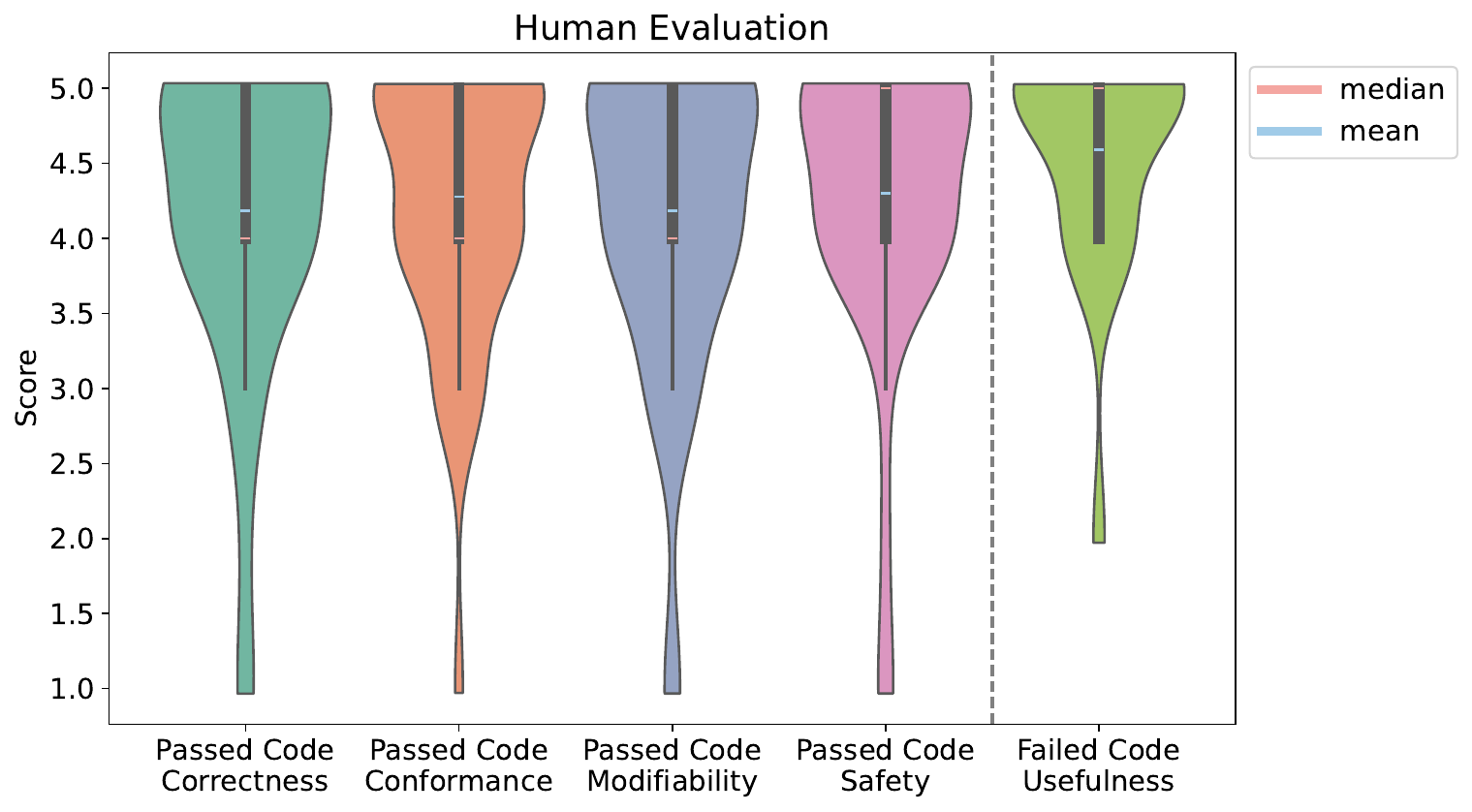}
    \caption{Violin plot showing the manual evaluation results.}
    \label{fig:violin}
    \vspace{-0.4cm}
\end{figure}

Ultimately, we collected 65 annotations(due to limited expert availability) : 25 for OSCAT (15 passing/10 failing), 17 for LGF (12 passing/5 failing), and 23 Competition entries (14 passing/9 failing). Figure \ref{fig:violin} depicts the distributions of scores, including min, max, median, and mean values for correctness, conformance, modifiability, and safety among passing cases, alongside usefulness scores for failing cases.

For the \emph{passed samples}, the average ratings for the four concerns across the three datasets are 4.18, 4.28, 4.19, and 4.53, respectively. Notably, the median score for safety reaches the maximum of 5.0. These scores are particularly impressive, especially considering that our participants grounded their assessments in practical applicability. They offered numerous constructive critiques and suggestions for the generated code. For instance, the engineer with seven years of experience, who adheres to a stringent standard for industrial control coding quality, determined that seven out of ten samples satisfied the stipulated requirements. Nevertheless, he highlighted several issues from various perspectives. For example, he flagged a potential safety concern: ``\emph{In industrial environments, the validity of input data must be taken into consideration. An error or default value should be returned when the frequency is zero or negative to prevent runtime errors that could compromise system safety.''} Other participants provided expert feedback on the code as well. 
One participant identified a Modifiability issue: ``\emph{LLMs may not be familiar with the functionality of the Find function. It can return the position within the string by itself; there is no need to use a for loop to search through the string.}''
One correctness issue is about the lack of domain knowledge of industrial control: \emph{``In manual mode, the interlock switch between forward and reverse rotation does not reflect the actual situation since the relay action is delayed; therefore, we cannot directly determine the status using the button's state.''}

We discerned that among the 41 cases that passed, four were appraised with \emph{medium correctness}. A closer inspection of the comments revealed that \emph{only one case was deemed incapable of fulfilling the requirements}. The remaining three raised some issues, such as intricate logic or unstable outcomes. Take, for instance, a task within LGF requiring the code that \emph{generates a random value of the real number type within a specified range}. While the participant confirmed the fundamental functional correctness, he also expressed reservations about \emph{the potential for the nanosecond component of system time to be recurrent or erratic in some PLCs due to disparate time resolutions. Consequently, it is recognized that our produced code may not guarantee entirely random values upon each invocation}.

Concerning \emph{conformance}, seven cases each exhibit one issue, with four related to \emph{naming conventions and commentary}, two associated with \emph{unnecessary use of extensive data structures}, and one case highlighting an \emph{improper mix of manual and automated procedures}.
Six cases were rated as having medium \emph{modifiability}, with the main points of contention being \emph{ambiguous naming} (three cases), \emph{unclear logical transitions} (two cases), and \emph{improper utilization of process variables}.
Lastly, five cases were considered to have medium \emph{safety} concerns, primarily due to \emph{inadequate input verification and subpar handling of exceptions or errors.} We highlight these cases because, in the context of industrial control code, non-functional quality attributes are crucial. Such aspects warrant greater focus and consideration from the community.
    
Despite containing some errors, the \emph{failed samples} received scores of 4.40, 4.30, and 4.89 within the three datasets, indicating that all participants regarded them as valuable for their coding tasks. Annotations from four participants consistently praised the samples' usefulness. For example, one participant observed, ``\emph{In industrial settings, it's common to encounter situations where slope processing is necessary to manage temperature rise and acceleration rates. ... The program also includes an error handling mechanism... While there is room for code Modifiability improvements, its complete and clear functionality means it can still be deployed as-is, offering significant gains in time efficiency.}''

\vspace{1mm}
\begin{custommdframed}
\textbf{Answer to RQ3:} \emph{AutoPLC} generated code that was positively evaluated by experts on selected datasets, though some cases revealed issues such as insufficient input validation or incomplete error handling.
\end{custommdframed}
\vspace{-0.2cm}

 \section{Discussion}
\label{sec:discussion}

\subsection{Error Types Analysis}
 
\begin{table}[ht]
\centering
\setlength{\abovecaptionskip}{0.1cm}
\scriptsize
\caption{Error types analysis of AutoPLC and Claude.}
\begin{tabular}{l|p{0.5cm}p{0.5cm}|p{0.5cm}p{0.5cm}|p{0.5cm}p{0.5cm}}
\toprule
\multirow{2}{*}{\textbf{Error Types}} & \multicolumn{2}{c|}{\textbf{OSCAT}} & \multicolumn{2}{c|}{\textbf{LGF}} & \multicolumn{2}{c}{\textbf{Competition}} \\ \cmidrule{2-7} 
&  \textbf{Claude} & \textbf{Ours} &  \textbf{Claude} & \textbf{Ours} & \textbf{Claude} & \textbf{Ours} \\ \hline
Undefined error  & 1.84 & 0.50  & 5.88 & 0.77  & 3.00 & 0.56 \\
Mismatch error  & 3.39 & 0.66  & 0.97 & 0.006  & 1.09 & 0.02 \\ 
Call error  & 0.05 & 0.004  & 2.12 & 0.01 &  2.00 & 0.27 \\
Type conversion error & 0.25 & 0.18  & 0.54 & 0.25  & 0.49 & 0.20 \\
Others & 0.21 &0.20  & 0.17 & 0.10  & 0.04 & 0.22 \\ 
\bottomrule
 
\end{tabular}
\label{tab:errorTypes}
\vspace{-0.3cm}
\end{table}
 
Through systematic analysis of frequent error patterns, we identify current limitations of \emph{AutoPLC} and outline pathways for future enhancement.
Specifically, to evaluate the improvements introduced by our proposed components, we systematically categorize error types from both \emph{AutoPLC} and its backbone model, Claude, into five distinct classes based on compiler diagnostics:
undefined errors, mismatch errors, call errors, type conversion errors, and other errors. Table~\ref{tab:errorTypes} reports the average error count of each type across benchmarks, 
with color highlights for visual comparison.
 
\emph{Undefined errors} were the most frequent error type in a majority of the benchmarks. While often triggered by referencing undeclared variables, they commonly stem from invalid definitions in the variable declaration section—\textit{e.g.}, illegal initializations—that cause variables to fail to be declared successfully, leading to downstream failures. Our staged repair mechanism mitigates this by prioritizing declaration fixes, reducing undefined error rates by 
72.8–86.9\% (vs. Claude).

\emph{Mismatch errors} (e.g., unmatched parentheses, symbols) were also prevalent. These were effectively reduced by retrieving cases from the same ST variant, which avoids structural hallucinations. 
We observed a reduction of 80.5–99.4\% compared to Claude.

\emph{Call errors} were particularly common in the LGF and Competition benchmarks, and typically stemmed from incorrect instruction usage—such as referencing non-existent instructions or mismatching parameters. These errors reflect limited model familiarity with vendor-specific instructions. AutoPLC mitigates this via instruction recommendation and full signature integration.

\emph{Type conversion errors} arise when argument types are not properly cast or when invalid conversion syntax is used—e.g., \textbf{\texttt{TO\_REAL}} instead of \textbf{\texttt{REAL\_TO\_REAL}}. These are frequent in ST due to its stricter type constraints. Our method explicitly encodes type expectations during code generation and applies compiler feedback to resolve mismatches.

\emph{Other errors}, although less frequent, such as invalid user types, redefinition, invalid variable, and unused label, were also addressed, which contributed to improving the overall compilation pass rate.

\subsection{Threats to Validity}
\label{subsec:threats}

\noindent \textbf{Internal Validity.} 
While Rq2ST explicitly excludes requirements exhibiting literal substring matches with benchmark cases, we acknowledge the potential existence of semantically similar examples. This characteristic intentionally reflects the practical reality of case repositories in code generation systems~\cite{MapCoder,RAGforCodeGen,liu2024agents4plc}, where semantic overlaps naturally occur. 

\noindent \textbf{Construct Validity.} While standard quantitative metrics (Compilation Pass Rate, Average Error Count)~\cite{fakih2024llm4plc,liu2024agents4plc} provide initial validation, they are insufficient for guaranteeing functional correctness. To address the lack of publicly available test benchmarks, we implemented an industry-peer-reviewed automated test case generation framework with rigorous quality filtering. Our experimental results demonstrate the method's efficacy through high pass rates (76.61\%).  

\noindent \textbf{External Validity.} While our framework can theoretically support generating ST variants beyond Siemens' SCL and CODESYS' ST through vendor-specific configurations of Rq2ST and APILib, the generalization performance requires further investigation and empirical validation.

\subsection{Limitations}
\label{subsec:limitations}

\noindent \textbf{Limited Semantic Checks.} Our current approach primarily corrects code based on compilation feedback, which may miss runtime issues. Further improvements could incorporate time-sensitive, and safety-critical autonomous test suites or simulations—for instance, validating deterministic behavior, resource constraints, hardware configurations, and I/O mappings.

\noindent \textbf{Limited Assessment of Functional Correctness, Safety, and Reliability.} Throughout our evaluation, we concentrate on metrics such as the Compilation Pass Rate and the Average Error Count during quantitative analysis (pertinent to RQ1 and RQ2). In addressing RQ3, annotators evaluate the generated code's 
correctness, conformance, modifiability, safety, and usefulness.
However, our evaluation does not include a thorough assessment of functional correctness, safety, and reliability using more formal and rigorous methodologies.

\noindent \textbf{Complex control logic and large-scale integration remain unexplored.} The current benchmark tasks primarily focus on fundamental programming scenarios, while intricate control logics (e.g., interlocking mechanisms, recipe management) and large-scale system integration require further investigation.

\section{Related Work}
\label{sec:relatedWork}

Automating program generation from natural language (NL) requirements has long been a research objective. Early approaches to PLC program generation relied on formal methods, rule-based systems, and manually defined features~\cite{steinegger2012automated, darvas2016PLC}. While effective in constrained domains, these methods required substantial human expertise and lacked adaptability to real-world industrial scenarios.

Recent advances in large language models (LLMs)~\cite{chatgpt, roziere2023codellama, guo2024deepseek} have shown promising code generation capabilities, with growing applications in industrial domains. Due to limited domain-specific training resources, most studies employ general-purpose LLMs directly. For instance, ChatGPT has been tested for IEC 61131-3 ST generation~\cite{ChatgptForPLC}, while \citet{koziolekLLMbasedControlCode2024} developed a P\&ID-to-ST translation approach. However, these methods suffer from low determinism and poor compilation rates; our experiments show GPT-4o achieves only 35.47\% success on ST generation tasks. Recently, a Siemens research group achieved 70\% compilation success by first pretraining LLMs on OSCAT libraries and then refining them with generated high-quality samples, as validated on both OSCAT and the APPS Python benchmark.

To address these limitations, recent work has focused on improving generated code quality through human-AI collaboration and retrieval augmentation. LLM4PLC~\cite{fakih2024llm4plc} introduced a user-guided iterative pipeline incorporating external verification, though its effectiveness varies with human expertise. Other approaches integrate RAG; \citet{koziolekLLMbasedRetrievalAugmentedControl2024} leveraged proprietary function blocks, while Agents4PLC~\cite{liu2024agents4plc} proposed a multi-agent framework combining LLMs with planning, RAG, and formal verification agents.

Key challenges remain in current methodologies: (1) the scarcity of high-quality, publicly available domain knowledge limits retrieval precision~\cite{koziolekLLMbasedRetrievalAugmentedControl2024, liu2024agents4plc}; (2) most systems target only specific ST variants, restricting generalizability; and (3) the absence of standardized benchmarks impedes objective comparison—existing evaluations range from case studies~\cite{koziolekLLMbasedControlCode2024} to small-scale task sets (40 tasks in~\citet{fakih2024llm4plc}, 23 in~\citet{liu2024agents4plc}).
\emph{AutoPLC addresses these by enabling vendor-aware generation, scaling evaluation to 945 tasks across two variants, and open-sourcing our domain knowledge base to advance the field.}

\section{Conclusion}
\label{sec:conclusion}

Industrial Structured Text development is often hampered by vendor dialect challenges. To address this issue, we introduced AutoPLC, an end-to-end framework for generating vendor-aware ST from NL requirement. AutoPLC integrates curated knowledge bases, task planning, retrieval-augmented generation, and industrial compiler-feedback improvement into a unified workflow.
Our experiments on CODESYS' ST and Siemens' SCL datasets demonstrated that AutoPLC  outperformed state-of-the-art baselines, particularly in compilation pass rates and error reduction. The importance of the core designs has been confirmed through ablation studies. Five ST experienced experts affirmed the usefulness of our generated code. In the future, we aim to tackle more complex logic and multi-device interfaces, and test on a broader set of real-world benchmarks derived from authentic engineering practices.

\bibliography{Reference}

\end{document}